# Could Planck level physics be driving classical macroscopic physics through a random walk?

## C. L. Herzenberg

**Abstract**
We examine a very simple conceptual model of stochastic behavior based on a random walk process in velocity space. For objects engaged in classical non-relativistic velocities, this leads under asymmetric conditions to acceleration processes that resemble the behavior of objects subject to Newton's second law, and in three dimensional space, acceleration dependent on an inverse square law emerges. Thus, a non-relativistic random walk would appear to be capable of describing certain prominent features of classical physics; however, this classical behavior appears to be able to take place only for objects with mass exceeding a threshold value which we identify with the Planck mass. Under these circumstances, stochastic space-time displacements would be smaller than the Planck length and the Planck time so that such classically behaved objects would be effectively localized. Lower mass objects exhibit more rapid diffusion and less localization, and a relativistic random walk would seem to be required of objects having masses comparable to and smaller than the threshold mass value. Results suggest the possibility of an intrinsic quantum-classical transition in the microgram mass range.

**Introduction - where we are going with this**

Random walk models have been used successfully in physics, chemistry, biology, economics and a number of other fields to address a variety of different problems.[1,2,3] Random walks serve as a fundamental model for stochastic or nondeterministic activity.

Many applications of random walks take place in ordinary physical space, but we will examine a very simplified model of a random walk of an object in velocity space. When we examine an asymmetric random walk (that is, a random walk with a preferred probability of steps along a particular direction) in velocity space, we find that it leads to an average acceleration process that bears a resemblance to Newton's second law. For random walks in three dimensional space, we express step probabilities in terms of solid angles, and find that dependence on solid angle leads to an inverse square law dependence for forces originating from discrete objects located at a sufficient distance from the stochastic object.

We go on to examine the mass dependence and magnitude of the stochastic parameters that describe random walks in velocity space with particular attention to gravitational and electrostatic accelerations. We find that classical random walks in this velocity space model can take place at non-relativistic speeds only for masses exceeding a threshold or reference mass, which we identify as comparable to the Planck mass. The displacements in space-time associated with such classical non-relativistic random walks seem to be restricted to extremely small values in the sub-Planck domain. This would appear to suggest that interactions in the Planck-level domain may be responsible for some of the prominent aspects of macroscopic classical behavior. These results of this model suggest that lower mass objects must undergo stochastic motion at relativistic speeds, and point to the existence of an intrinsic quantum-classical transition in a mass range overlapping the Planck mass.

We will use a very simplified approach throughout this study with the objective of developing insight and identifying some results of possible interest. This paper is largely limited to an inductive approach, applying qualitative and semi-quantitative arguments, considerations, and justifications, with the hope that a similar but more rigorous treatment can be developed in future work. Note that because of the semi-quantitative nature of this study, for the most part the equations should be regarded as only approximate estimates.

**The random walk**

A random walk is a mathematical formalization of a trajectory that consists of taking successive random steps.[1,2,3] In the simplest random walk models, it is assumed that a particle or object moves in steps (that is, one dimensional displacements in a space), taking steps of equal length in any available direction during equal discrete intervals of time.[1] In the case that steps in any direction occur with equal probability (which defines a symmetric random walk), over time the particle undergoes diffusion with no net systematic drift. (If in addition the particle is subjected to an externally applied force, it will acquire a drift velocity which is determined by the direction and magnitude of the applied force, the step time, and the mass of the particle.[2] However, alternatively, random walks with differing probabilities for steps along different directions are also possible; these are referred to as asymmetric random walks and these can also lead to a net drift velocity.[1]

If the random walk is limited to one dimension, we can designate the probabilities associated with the forward and backward directions as p and q. Then if $p = q = ½$, the random walk is symmetric, and there would be no net drift motion in either direction.[1] If $p = 1$ and $q = 0$, then the particle would be expected to move forward toward the right at the maximum possible speed; while if $p = 0$ and $q = 1$, the particle would be moving backward to the left at the maximum possible speed. At intermediate values for the probabilities, the object would undergo drift at intermediate speeds.

**Random walk in velocity space**

So far we have provided only a minimalist description of a random walk in ordinary physical space (also referred to as position space or real space or laboratory space or configuration space or coordinate space). But it is also possible to describe and examine physical processes in other types of 'spaces'. A 'velocity space' is a mathematical space in which each point in the 'space' corresponds to a particular velocity value; the coordinates of each point are the velocity components along each of the axes which correspond to directions in ordinary position space.[4,5] We will be concerned primarily with investigating a random walk in velocity space.

Accordingly, instead of concentrating on a model in which a particle moves a step to the right or left in ordinary space for every step in time, we will start by considering a model in which a particle moves to the right or left in a one-dimensional velocity space by a fixed amount, that it, it changes its velocity by a fixed amount for every step in time. (This of course is another simplification; in a more realistic treatment of a physical problem we might expect to be dealing with some sort of average values of the variables.)

We can write the velocity in a random walk model as:

$$v_{n+1} = v_n + \Delta v \qquad (1)$$

where in this equation the step in velocity $\Delta v$ can be positive or negative and the integer n designates a step number.

In this one dimensional velocity space, this operation would correspond to increasing or decreasing the object's velocity in stepwise fashion, so that its velocity could take on discrete values of an initial speed plus or minus integral multiples of a fixed amount of speed (the step speed).

Let's consider a random walk in one dimensional velocity space in more detail. Velocity steps of magnitude $\Delta v$ to the right or left are taken at time intervals $\Delta t$ and with probabilities p or q (to the right or left respectively; and, since no other directional options are available in one dimensional space, p + q = 1.) Each step in velocity space involves a change in velocity with time (that is, formally, an acceleration) of net magnitude $a_u = \Delta v/\Delta t$ toward the right or toward the left. If p = q, then we can expect, on average, equal numbers of step changes in velocity to the right and to the left, averaging to a net average value of zero change in velocity (that is, zero net average acceleration); and thus the average velocity value will remain at the initial conditions. This would correspond to a symmetric random walk in velocity space. Thus, a symmetric random walk would describe diffusion in velocity space (associated with an uncertainty in velocity) without systematic drift. However, if p > q, then the particle will, on average, exhibit drift in velocity space and accelerate on average toward the right; whereas if p < q, then the particle will accelerate on average toward the left.

Accordingly, the quantity (p - q) can be expected to provide a measure of the probability of an average acceleration toward the right, while (p + q) will correspond to the sum of the probabilities of motion in either available direction, and thus be equal to unity. Consequently, the fraction of the time $f_t$ on average that the particle would spend accelerating toward the right rather than engaging in random motion in velocity space might then be expected to be given by $f_t = (p - q)/(p + q)$.

It would appear then that the net average acceleration a of the particle over time would therefore be expected to be given by:

$$a = a_u f_t = (\Delta v/\Delta t)f_t = (\Delta v/\Delta t)[(p - q)/(p + q)] \qquad (2)$$

Thus, we see that the net average acceleration of the particle is given by the product of the magnitude of a net elementary acceleration experienced by the particle during any single step in velocity space (a quantity that is intrinsic to the stochastic motion) with a quantity in brackets that is a measure of the probability that excess steps will be taken along a particular direction. More generally, we can regard the net average acceleration as the product of an elementary acceleration per step times a ratio of probabilities (e.g., the ratio of the probability of a step toward the right to the probability of a step in any direction).

While the velocity space has the most basic role in this approach, we also need to consider an associated ordinary position space in conjunction with the velocity space, since the physical object of interest ultimately resides in ordinary space. This ordinary position space would also be one dimensional, with direction definitions in correspondence or conformity with those of the associated velocity space, and in principle a mapping could be made between the motion in ordinary position space and the particle's motion in velocity space.

If we wish to interpret the preceding equation, Eqn. (2), in terms of classical physics, we might compare it with Newton's second law F = ma, where F is the applied force, m is the mass, and a is the resultant acceleration. We can accordingly view the quantity a which is evaluated as an acceleration in Eqn. (2) as representing a force-to-mass ratio:

$$F/m = a = a_u[(p - q)/(p + q)] = (\Delta v/\Delta t)[(p - q)/(p + q)] \qquad (3)$$

Comparing the numerators and denominators in Eqn. (3) suggests that a force acting on such a stochastic object would relate to the difference between the probabilities for the object to engage in steps in velocity to the right or left (and thus to the excess number of steps taken along a particular direction), while the mass associated with the object would relate to the total number of steps in velocity taken in any direction. Thus we see that even in its very simplest manifestation, a random walk in velocity space can lead to behavior somewhat resembling basic classical behavior governed by Newton's law, and the force and mass parameters would have related quantities in the random walk model.

(Some care is needed in making and interpreting such a comparison, as Newton's law describes behavior in ordinary position space.)

**Random walk in three dimensions**

But what would such a random walk look like in a higher dimensional space? In one dimension, we only need to consider the probabilities of steps to the right and to the left.

But in higher dimensions, we would need to consider steps occurring in all possible directions and their associated probabilities.

In three dimensional velocity space, the multitude of directions available for a step means that during almost all of the time, steps in a random walk would go along different directions, and only infrequently would steps move along precisely the same direction. Thus, changes in direction would take place frequently, while changes in speed would take place only rarely. Therefore, as a good approximation, we can treat the steps in velocity space as generally taking place along different directions while being characterized approximately by an average step speed that we will designate by $v_{avg}$.

Furthermore, since in higher dimensions, we need to consider steps occurring in all possible directions and their associated probabilities, we will have directions together with their associated magnitudes to consider. If we join together a direction and an associated magnitude, we have what amounts mathematically to a vector which would describe the probability of acceleration in a particular direction. Thus, we could anticipate that entities causing or leading to acceleration in this model would be described by vectors.

In the case of a three dimensional space, the direction of the instantaneous velocity could point along any direction in three dimensions, not just along a particular axis. If changes in velocity direction are random, presumably we will be dealing with a symmetric random walk in three dimensional velocity space. But if the random walk is not symmetric, having a net excess along one direction could be expected to lead to an increasing average velocity along the direction of asymmetry, that is, a net acceleration along the direction of asymmetry. Thus, as noted, we can associate accelerations with directional solid angles in three dimensional space.

As before, we can approach this problem in further detail by segmenting it into two parts: First, a symmetric random walk in three dimensions (that is, equal probabilities associated with motion along every direction in three dimensional velocity space). Except along a particular direction, say, along the $v_x$-axis, along which an excess probability would be present that would lead to a net increase in velocity, that is, an acceleration of the object along the $v_x$ axis.

As noted earlier, we will consider the velocity space of primary interest, but we will also wish to consider a correlated ordinary position space. Thus, if we are dealing with a three dimensional velocity space, such that velocities can be identified as velocity space positions characterized by components along the axes such as $v_x$, $v_y$, and $v_z$; then we will also introduce an associated real position space where spatial locations can be characterized by associated spatial coordinates x, y, and z. So, for example, the $v_x$ axis and the x axis could be considered in a practical sense aligned. Thus, a velocity or acceleration aligned along the $v_x$ axis (for example) in velocity space would be aligned along the corresponding x axis in the associated position space.

Accordingly, in analogy with Eqn. (2) or Eqn. (3), we might write for the acceleration in three dimensional space:

$$a = a_u(p_{dir}/p_{all}) = (\Delta v/\Delta t)(p_{dir}/p_{all}) \qquad (4)$$

Here, $p_{dir}$ is the probability of excess velocity steps along a particular direction in velocity space, and $p_{all}$ is the probability of velocity steps in all possible directions in velocity space.

If we introduce an average speed $v_{avg}$ that characterized the stepping motion as an estimate of the step size in velocity space, Eqn. (4) may be reexpressed approximately as:

$$a = a_u(p_{dir}/p_{all}) = (v_{avg}/\Delta t)(p_{dir}/p_{all}) \qquad (5)$$

Thus, the net acceleration is estimated as the average speed divided by the step time, times the ratio of the probability of stepping in the direction of acceleration to the probability of stepping in any direction at all, or in all other directions.

**Spherical coordinates in three dimensional space**

In dealing with an object in three dimensional space, it is for many purposes more helpful to use spherical polar coordinates than rectangular Cartesian coordinates.

Thus we might prefer to use spherical polar coordinate systems to describe both the ordinary space associated with a velocity space as well as the velocity space itself. We could regard the velocity space spherical polar coordinate system as centered on zero velocity or an initial velocity, while we could regard the associated coordinate space spherical polar coordinate system as centered on an initial location of or near the object under consideration.

As already noted, results so far suggest that entities causing or leading to acceleration in this model would be described by vectors. Or, alternatively, entities leading to acceleration could instead be described by alternative realizations of directions and associated magnitudes, such as by using directions together with the magnitudes of the associated opening angles, or directional solid angles in three dimensional space.

We can again approach the problem by partitioning it into two parts: First, a symmetric random walk in three dimensional velocity space (that is, equal probabilities of moving along every direction in three dimensional velocity space). Then, to allow for an asymmetric random walk, along one selected direction we will describe a small solid angle within which there would be an excess probability of velocity steps (this would correspond to the quantity $p_{dir}$ in the preceding discussion). But having a net excess probability of steps along one direction would presumably lead to an increasing average velocity along the direction of asymmetry, that is, a net acceleration along the direction of asymmetry, in this case, along the direction of the solid angle. Thus we can associate accelerations with directional solid angles in three dimensional space. Accordingly, the particle undergoing random walk will not only diffuse in all directions, but will also undergo drift in velocity and thus accelerate on average along the axis of that particular solid angle characterized by a excess probability.

How large will the acceleration along the axis of that solid angle be? The particle is stepping about in all directions in three dimensional velocity space, and so we can presume from the otherwise overall uniformity of that motion on a time-averaged basis that the fraction of time allocated to motion in that special solid angle would on average be proportional to the magnitude of the solid angle. If we designate the particular solid angle $\Omega_v$ in velocity space, and take the ratio of that solid angle in steradians to the full solid angle encompassing all directions in three dimensional space (which would be $4\pi$ steradians), then we would find for the fraction $f_v$ of the full solid angle that $f_v = \Omega_v/4\pi$. Correspondingly, we would expect the fraction of time that the particle is engaged in acceleration in that selected direction also to be approximated by $f = \Omega_v/4\pi$, while the rest of the time it would be moving to and fro aimlessly in random motion along all other possible directions of velocity space.

As a result, we might expect the net average acceleration of such a stochastic particle over time to be given (for sufficiently small solid angles) in three dimensional velocity space by:

$$a = a_u(\Omega_v/4\pi) = (v_{avg}/\Delta t)(\Omega_v/4\pi) \tag{6}$$

In this expression, all of the parameters refer to values in velocity space.

Again we see that the acceleration of the particle is given by the product of the magnitude of an intrinsic acceleration associated with the random walk process of the particle in velocity space, with a quantity that is a measure of the effect of the outside world in affecting the likelihood of the random walk being distorted so as to provide for preferential step motion along a particular direction.

**Solid angle and inverse square law force behavior**

Next, it would be useful to be able to develop an expression analogous to Eqn. (6) in association with three dimensional ordinary position space. As its velocity changes in a random walk, a stochastic object will also be changing positions rather randomly in ordinary position space, while also on average accelerating along the particular direction that corresponds to the direction of acceleration in velocity space. Accordingly, the object undergoing random walk will only diffuse in all other directions, but will also on average accelerate along the axis of an associated solid angle in ordinary position space.

If the random walk in velocity space takes place at non-relativistic velocities, the time intervals and the average speed value and other parameters would be expected to be in correspondence with those measured in the associated position space, as no relativistic effects from high speed motion would need to be taken into account. Thus, essentially the same equation as Eqn. (6) with similar parameters might be expected to apply in ordinary position space. An associated solid angle $\Omega_s$ would need to be defined in ordinary space, and to conserve the magnitude of the effect, we would have to require $\Omega_s = \Omega_v$. Thus, from Eqn. (6), we can write:

$$a = a_u(\Omega_s/4\pi) = (v_{avg}/\Delta t)(\Omega_s/4\pi) \qquad (7)$$

If the entity responsible for affecting the directional probabilities of steps taken in different directions by the object undergoing random walk is another identifiable localized object (such as a discrete object in three dimensional space), then, in ordinary position space, the causative object would be located at a specifiable distance from the object in stochastic motion. We have already identified the direction associated with the influencing object as relevant; if its distance in position space might be of relevance also, then it would be appropriate to describe its influence in terms of a solid angle in ordinary position space, even if physics in velocity space is fundamentally involved. But then we can describe a solid angle $\Omega_s$ in ordinary space in terms of an area A associated with the distant object, together with its distance r from the responding object, as:

$$\Omega_s = A/r^2 \qquad (8)$$

Then, combining Eqn. (7) with Eqn. (8), we can write for the acceleration to be expected from this model:

$$a = a_u(\Omega_s/4\pi) = (v_{avg}/\Delta t)(A/4\pi r^2) \qquad (9)$$

Here, the quantities $v_{avg}$ and $\Delta t$ characterize the object undergoing stochastic motion, whereas the quantity A characterizes the other object that is affecting the directional probabilities of motion of the stochastic object and the characteristics of the interaction. Thus, we might expect that the parameters $v_{avg}$ and/or $\Delta t$ would in some way incorporate the inertial mass of the stochastic object, while the parameter A would incorporate relevant properties of the other object that is affecting the motion of the stochastic object, as well as some properties of the stochastic object.

Thus, we see that this simple random walk model in velocity space appears to lead to the expectation of inverse square law ($1/r^2$) behavior to characterize forces between separated objects at a distance from each other in a three dimensional world. Accordingly, this approach would appear to be suggestive for modeling both gravitational and electrostatic accelerations. However, the description of other types of force behavior might be less amenable to analysis, particularly for separation distances comparable to the distances associated with steps of the random walk, so that other forces acting on stochastic objects might exhibit different dependences on distance.

**Electrostatic acceleration**

Next let us look at the application to a specific case, the classical electrostatic force and resultant acceleration.

In the case of the electrostatic force, we would ascribe an increase or reduction in probability of steps in velocity space along a particular direction to the presence of an electrically charged object located in the corresponding direction in ordinary position space.

Let us compare the expression for the acceleration of an object undergoing an asymmetric random walk with the acceleration caused by an electrostatic force between two separated point objects. The equation describing electrostatic acceleration $a_q$ of an object of mass m would be:

$$a_q = (1/m)(q_1 q_2 / 4\pi\varepsilon_o r^2) \tag{10}$$

Here, $q_1$ and $q_2$ are the electrical charges associated with the two objects, $\varepsilon_o$ is the dielectric constant of free space, and m is the mass of the stochastic object whose acceleration is being described. For simplicity we are assuming that the attracting object is sufficiently massive so that center of mass effects can be neglected; then r is the distance separating the objects.

If we equate Eqn. (9) for the acceleration of a stochastic object to the classical electrostatic acceleration given by Eqn. (10) we obtain for the electrostatic acceleration $a_q$:

$$a_q = (1/m)(q_1 q_2 / 4\pi\varepsilon_o r^2) = a_u(A_q/4\pi r^2) = (v_{avg}/\Delta t)(A_q/4\pi r^2) \tag{11}$$

Here, $a_u$ is the step acceleration and $A_q$ would be a parameter with dimensions of area that characterizes the electrostatic interaction effect on the stochastic motion.

It can be seen that this model seems to be capable of reproducing some general features of classical electrostatic acceleration. Here, it would appear that the quantities $v_{avg}$ and $\Delta t$ characterize the object undergoing the acceleration, whereas the quantity $A_q$ includes properties of the object creating the electrostatic field and the interaction, and will include properties of the stochastic object as well. We can anticipate that the area parameter $A_q$ will incorporate the magnitude of the electrical charges and the electrical conductivity of free space.

**Gravitational acceleration**

Next let us look at the application to another specific case, classical gravitational attraction and acceleration.

In the case of classical gravitational attraction, we would ascribe an increase in the probability of steps in velocity space along a particular direction to the presence of a gravitating object located in the corresponding direction in ordinary position space.

Let us compare the expression for the acceleration of an object undergoing an asymmetric random walk in velocity space with the acceleration caused by a gravitational force between two separated point objects. The equation describing gravitational acceleration $a_G$ of an object of mass m would be:

$$a_G = (1/m)(GmM/r^2) \tag{12}$$

Here, G is the gravitational force constant, and m and M are the masses of the two objects, with m being the mass of the stochastic object whose acceleration is being analyzed. For simplicity we are assuming that the attracting object of mass M is sufficiently massive so that center of mass effects can be neglected; then r is the distance separating the objects.

If we equate Eqn. (9) for the acceleration of a stochastic object to the classical gravitational acceleration given by Eqn. (12) we obtain:

$$a_G = (1/m)(GmM/r^2) = a_u(A_G/4\pi r^2) = (v_{avg}/\Delta t)(A_G/4\pi r^2) \tag{13}$$

Here, $a_u$ is the step acceleration and $A_G$ would be a parameter with dimensions of area that characterizes the gravitational interaction effect on the stochastic motion.

It can be seen that this model reproduces some general features of classical gravitational acceleration. Here, it would appear that the quantities $v_{avg}$ and $\Delta t$ characterize the object undergoing the acceleration, whereas the quantity $A_G$ includes properties characterizing the object creating the gravitational field and the interaction, and will include properties of the stochastic object as well. We can anticipate that the area parameter $A_G$ will incorporate the magnitudes of the gravitational masses of the two objects.

**Mass dependence and magnitude of stochastic parameters**

In order to be able to understand the predictions of this type of random walk model and compare them with known expressions for acceleration from familiar forces, we will examine the influence of the various parameters that enter the equations, and attempt to establish some values for these parameters. This will include estimating the magnitudes of the steps in time, the steps in velocity and the relevant area parameters. In order to begin to address this in the simplest possible manner, we will backtrack temporarily to one dimension.

For a random walk in velocity space, Eqn. (3) gives the acceleration (or the force-to-mass ratio classically) as a product of two quantities: an elementary acceleration $a_u = (\Delta v/\Delta t)$ characterizing the random walk, and a dimensionless ratio of probabilities (the ratio of the probability associated with stochastic motion toward a particular direction compared to the sum of the probabilities of stochastic motion toward all available directions, - that is, the ratio (p – q)/(p + q)).

As noted earlier, the relationship for the force-to-mass ratio given in Eqn. (3) suggests that a force acting on such a stochastic particle might be related to the difference between the probabilities for the particle to engage in steps in velocity to the right or to the left (and thus to the excess number of steps taken along a particular direction per unit time), while the inertial mass associated with the particle could be related to the total number of steps in velocity in any direction at all, per unit time. In particular, the mass would seem to be proportional to a frequency associated with the random walk (that is, the frequency of the steps taken in the random walk). Thus, we find an indication of a proportionality between the mass of an object and a frequency as the number of steps per unit time associated with its stochastic motion; so we might anticipate a relationship of the form f ~ m or f = km, where k is a constant factor.

Even while we are looking at a classical case, we can also note a resemblance to quantum behavior in the suggestion of a connection between the mass of an object and an associated frequency. In quantum mechanics, a proportionality exists between frequency and mass in the de Broglie frequency relationship $f = mc^2/h$ which associates a quantum frequency f to a mass m; and where h is Planck's constant and c is the speed of light.

Returning to the perspective of classically behaved stochastic objects, we have just seen that the frequency of steps in the random walk would seem to be proportional to the mass. And, accordingly, the inverse of the stochastic frequency which would correspond to the step time Δt would thus be expected to be inversely proportional to the mass, that is, Δt ~ 1/m. Let us make this inverse relationship between the step time and the mass explicit in the form:

$$\Delta t = t_R(m_R/m) \qquad (14)$$

Here, $t_R$ and $m_R$ are simply constants having the dimensions of time and mass respectively which have been introduced to formalize the proportionality and provide for the proper dimensions in the equation. We will refer to the constants $t_R$ and $m_R$ as the reference time and the reference mass respectively. Eqn. (14) makes it explicit that the step time is expected to be inversely proportional to the mass in this approach. Thus, we can anticipate that the step times would be small for large masses, and also the step times would be expected to be large for small masses.

Let us now consider how to approach the evaluation of the other parameters.

Equating the earlier equations for gravitational acceleration and for electrostatic acceleration to the corresponding classical expressions for these accelerations, we obtained Eqn. (11) and Eqn. (13) which provided effectively only two equations for four unknowns, the quantities $v_{avg}$, Δt, $A_G$ and $A_Q$. Thus, we cannot solve for the parameters directly. However, we could attempt a first pass at getting closer to evaluating these quantities by comparing the foregoing equations and partitioning the parameters and allocating common quantities in the two equations to the common elementary acceleration ratio $a_u = (v_{avg}/\Delta t)$ that describes the stochastic motion and differing quantities in the two equations to the parameter A (which describes the interaction, and would differ, depending on the force, and correspond to $A_G$ or $A_Q$ respectively). Following that we can go about adjusting the dimensions of the quantities that we obtain with the help of relevant physical constants.

So, as a trial procedure, let us do this next, and allocate the differing quantities associated with the specific forces to the area parameters, while provisionally allocating the (1/m) factor (which seems common to both cases) to the parameters dependent on the random walk itself, namely to the elementary acceleration associated with each step, $a_u = v_{avg}/\Delta t$.

If this partitioning of factors is correct, then the elementary acceleration, which is given by $v_{avg}/(\Delta t)$, would be expected to vary inversely with the mass m of the object engaged in the random walk; that is, that $v_{avg}/\Delta t \sim 1/m$. And as a result of this inverse dependence on the mass, an object with a large mass would tend to have only small changes in its elementary acceleration during its steps in time, whereas objects having small masses would tend to have large changes in elementary acceleration during their steps in time.

But as discussed earlier, there is already evidence to indicate that the step time $\Delta t$ is inversely proportional to the mass, that is, $\Delta t \sim 1/m$. If both the step time $\Delta t$ and the elementary acceleration $v_{avg}/(\Delta t)$ vary inversely with the mass, we can conclude that the step velocity or average velocity must then vary inversely as the square of the mass, that is, that $v_{avg} \sim 1/m^2$. We can make this dependence explicit by expressing the ratio of the average velocity to the speed of light in terms of a ratio of masses, using the (still arbitrary) constant with the dimensions of mass that we designated as the reference mass, $m_R$. Accordingly, we can write:

$$v_{avg}/c = m_R^2/m^2 \qquad (15)$$

Eqn. (15) makes it explicit that the average stochastic velocity would be expected to vary inversely with the square of the mass of the object undergoing the random walk.

Now we can also write down an expression for the elementary step acceleration $a_u$ in terms of the constants already introduced as:

$$a_u = v_{avg}/\Delta t = c(m_R^2/m^2)(m/t_R m_R) = (c/t_R)(m_R/m) \qquad (16)$$

Now that we have arrived at expressions for three of the most important parameters describing the random walk, we can use them to explore other aspects of the stochastic motion.

We can now examine approximately how the size of the steps that occur in ordinary space may be expected to depend on the mass.

The magnitude of a displacement in distance $\ell$ in ordinary space associated with a step in velocity space would be given at least approximately by:

$$\ell = v_{avg}\Delta t \qquad (17)$$

We can evaluate this displacement $\ell$ given in Eqn. (17) by combining it with Eqn. (15) and with Eqn. (14) to obtain:

$$\ell = c(m_R^2/m^2)[t_R(m_R/m)] = ct_R(m_R^3/m^3) \qquad (18)$$

Thus, the step displacement in ordinary space would in this approximation be expected to vary approximately inversely as the cube of the mass.

Returning to the behavior of the step speed, what would be the implications of a step speed inversely proportional to the square of the mass, as indicated in Eqn. (15)?

Eqn. (15) would indicate that for very large masses, the step speed or average speed of stochastic motion would tend to be very slow; while for very small masses, the step speed or average speed of stochastic motion would tend to be very fast. Furthermore, for very large masses, the step times would be short (in accordance with Eqn. (14)). Thus, short step times and slow step speeds would be expected for high mass objects; while long step times and rapid step speeds would be expected for low mass objects. Accordingly, we might anticipate slower diffusion and greater localization for high mass objects and more rapid diffusion and less localization for low mass objects.

The analysis so far has been for random walks involving non-relativistic velocities. Thus the present non-relativistic considerations would be expected to apply only to the case of higher mass objects, which would be moving at slower speeds.

We have seen that this inverse dependence of stochastic speed on the square of the mass indicates that the stochastic motion of lower mass objects should take place at much higher speeds, including the possibility of extending into relativistic speeds. In fact, Eqn. (15) would require that the present classical analysis lead to a step speed reaching the speed of light at a mass equal to the reference mass $m_R$ and exceeding the speed of light in the range of masses below the reference mass $m_R$ (an unacceptable result indicating the need for a relativistic treatment of the problem in this lower mass range). Accordingly, Eqn. (15) could apply only to objects of mass appreciably larger than the constant reference mass $m_R$, so as to limit the step speeds to well below the speed of light. Thus, the constant reference mass $m_R$ sets a lower mass limit below which this analysis involving classical non-relativistic motion would not be applicable.

At a value of the mass equal to this limiting mass $m_R$, we would find from Eqn. (14) that the step time would take on the reference time of $t_R$. But what might that be? Since we are examining the lower mass limit of applicability of classical behavior, it would seem reasonable to try to employ the de Broglie frequency as a guide for estimation purposes.

If we associate a de Broglie frequency $mc^2/h$ with a mass m we might expect a time step of the order of $h/mc^2$; let us take that to be a tentative guide. Then in Eqn. (14), for the mass $m_R$, we might expect that the associated reference time $t_R$ could be given by the de Broglie frequency relationship:

$$t_R = h/m_R c^2 \tag{19}$$

Inserting Eqn. (19) into Eqn. (14), we find:

$$\Delta t = h/mc^2 \tag{20}$$

That is, we are led back to a general use of the de Broglie frequency relationship for connecting the step time to the mass of the object engaged in the random walk, with applicability throughout the range of classical non-relativistic random walks.

As noted earlier, when pushed beyond its limitations, the present classical analysis would suggest that for extremely small masses, the step speed would approach infinity, an unacceptable result that within the classical framework is simply suggestive of speeds approaching the speed of light. However, that would lead to objects with very small masses engaging in a relativistic random walk, contrary to our modeling so far of non-relativistic behavior. This would require that we redirect our attention to relativistic random walks for the cases of small mass objects. However, relativistic random walks are considerably more complicated to deal with.[6] Accordingly, such high speed motion will be considered more specifically in a later analysis dealing with relativistic stochastic motion.

From the above results, it would seem that, if we are to retain non-relativistic stochastic behavior, we must limit our consideration to large masses. Large compared to what?

We have already introduced a constant with the dimensions of mass, the reference mass $m_R$, whose value would provide a basis for such a comparison. Can we get an idea of the magnitude of the reference mass $m_R$, so that we can get a better understanding of both the implications of and the limitations of this approach?

So what is the mysterious quantity that plays the role of the reference mass, $m_R$? It is a mass value above which classical, non-relativistic physics can seemingly provide a description of an interesting random walk process, and below which this classical behavior seems to break down. It is a constant with the dimensions of mass which we might expect to be composed of other physical constants; and, judging by the parameters that appear in, for example, Eqn. (13) and Eqn. (15) and Eqn. (20), we might expect these physical constants to include G, c, and h. The physical constant with the dimensions of mass that presents itself immediately as involving these constituents and as a candidate for comparison is the Planck mass, which is the unit of mass in the natural system of units known as Planck units, and is defined as $m_P = (hc/2\pi G)^{1/2}$.[7]

Let us tentatively assume, in order to explore the consequences, that the reference mass is in fact equal to the Planck mass; that is, that $m_R = m_P$:

$$m_R = m_P = (hc/2\pi G)^{1/2} \tag{21}$$

First off, we note that, unlike the other Planck units that are extraordinarily small by almost any standards, the Planck mass has a value of about 20 micrograms, and it is actually in the mass range where it could provide a reasonable mass value separating manifestly classical behavior from clearly quantum behavior.

If the Planck mass has this important role, we would expect it to enter into all of the preceding equations involving the reference mass. In particular, if we rewrite Eqn. (15) in terms of the Planck mass, we obtain an explicit equation for the step speed:

$$v_{avg}/c = m_P^2/m^2 = hc/2\pi G m^2 \tag{22}$$

Earlier, we evaluated the step time given by Eqn. (14) in terms of the constant reference time $t_R$ given by Eqn. (19). Inserting Eqn. (21), we now see that $t_R$ can be expressed as:

$$t_R = h/m_R c^2 = h/m_P c^2 = (2\pi hG/c^5)^{1/2} \tag{23}$$

We can now also evaluate the estimate of the basic step acceleration $a_u$ from Eqn. (16) as:

$$a_u = v_{avg}/\Delta t = c^4/2\pi Gm \tag{24}$$

It seems interesting to note that the elementary acceleration $a_u$ is independent of Planck's constant.

Let us compare the reference time $t_R$ that we obtained in Eqn. (23) with the Planck time. The unit of time in the Planck units, the Planck time, is given by $t_P = (hG/2\pi c^5)^{1/2}$, is equal to about $5 \times 10^{-44}$ seconds, and is the time required for light to travel a distance of one Planck length.[7] We find that:

$$t_R = 2\pi t_P \tag{25}$$

Thus, we can recognize that the reference time $t_R$ is closely related to the Planck time $t_P$, differing only by a reasonably small numerical factor. Accordingly, it appears that if the reference mass is in fact the Planck mass, then of necessity the reference time will be close to the Planck time. Furthermore, using Eqn. (21) and Eqn. (25), we can reexpress Eqn. (14) to obtain an explicit evaluation of the step time in terms of Planck units as:

$$\Delta t = 2\pi t_P (m_P/m) \tag{26}$$

We have recognized that a classical stochastic random walk can only take place for objects of mass somewhat exceeding the reference mass (and thus exceeding the Planck mass) so as to avoid reaching or nominally exceeding the speed of light in this classical estimate. From Eqn. (26), we now see that for any mass much exceeding the Planck mass, that the step time will necessarily be shorter than the Planck time.

We can also express the step displacement in ordinary space approximately from Eqn. (18) by entering the values for the reference mass and the reference time Eqn. (21) and Eqn. (25); we can then express the estimated step displacement in terms of the Planck length $\ell_P$ which is equal to $(hG/2\pi c^3)^{1/2}$ and corresponds to the distance traversed by an object travelling at the speed of light during the Planck time.[7] We find as an estimate for the step displacement in ordinary space:

$$\ell = ct_R(m_R^3/m^3) = 2\pi\ell_P (m_P^3/m^3) = (h^2/2\pi G)(1/m^3) \qquad (27)$$

Thus, the step displacement would be given roughly by the Planck length for masses comparable to the Planck mass, and would be reduced rapidly at higher masses, dropping off as the cube of the mass in this approximation.

While as noted earlier the magnitude of the Planck length and the Planck time are extremely small quantities, the Planck mass has a value within the range of direct human experience, at about 20 micrograms.[7] The ordinary familiar objects in our world that behave classically are macroscopic objects, consisting to a significant extent of objects with masses in excess of this amount. So it appears that a classical, non-relativistic random walk that would describe ordinary objects would seem to take place in the realm of the Planck scale, and involve distances smaller than the Planck length and times shorter than the Planck time. Thus, such a stochastic random walk of ordinary, familiar classical objects would seem to take place below what has been referred to as the smallest meaningful size, in the realm of strong quantum fluctuations of space-time, a region that has been called quantum foam.[8] Thus when the steps of the random walk are comparable to the Planck length and the step times become comparable to the Planck time, it would appear that we are no longer dealing with a random walk in ordinary space-time.

Backtracking somewhat, we can now also evaluate some of the remaining stochastic parameters. In particular, we can also evaluate the area constants present in the equations describing acceleration by solving Eqn. (11) and Eqn. (13) and entering evaluations of the parameters. We find:

$$A_G = 4\pi GM \, \Delta t/v_{avg} = 2\pi G^2 mM/c^4 \qquad (28)$$

And:

$$A_Q = (\Delta t/v_{avg})(q_1 q_2/4\pi\varepsilon_o m) = (2\pi G/c^4)(q_1 q_2/4\pi\varepsilon_o) \qquad (29)$$

It is interesting that the areal parameter describing the electrostatic interaction incorporates the gravitational force constant as well as electrical parameters.

Because of cancellations, this approach would lead us back to classical acceleration with the familiar force laws. Accordingly, it would appear to be compatible with ordinary classical motion for the case of objects of sufficient mass.

For dealing with smaller masses, a problem seems to be that we have started too easy with a non-relativistic approach, while the physical situation would seem to be relativistic. We can do a quick check on that. The quantum frequency that characterizes the behavior of a quantum particle of mass m is the de Broglie frequency $f = h/mc^2$, while the characteristic distance associated with the intrinsic quantum behavior of a mass is its Compton wavelength $\lambda_c = h/mc$; taken together these point toward an effective stochastic speed at or near c. Thus, when we are dealing with quantum particles, it would seem that the stochastic behavior would necessarily be relativistic, and our initial efforts to obtain some useful information using non-relativistic physics must be improved on. That will be addressed in a later paper.

It may be noted that a separation between the familiar classical behavior of macroscopic objects and the quantum behavior of microscopic objects does seem to occur (for entire objects rather than their parts) in a mass range very roughly comparable in magnitude to the Planck mass, which has a value of approximately 20 micrograms. Thus, the present approach would seem to be compatible with an intrinsic quantum-classical transition in the observed range, although other physical effects might also lead to intrinsic quantum-classical transitions, and decoherence effects can extend transitions into a much lower mass range.[9, 10]

**Discussion**

We have been exploring some outcomes of random walks in velocity space: in this model, we have seen that, in the presence of excess directional probabilities, these random walks can lead to accelerations reminiscent of Newton's second law; that the circumstances that cause accelerations in random walks in three dimensional space seem to lead to obligatory inverse square law behavior for force laws operating between well separated objects; that non-relativistic random walks appear to describe behavior of classical objects, while relativistic random walks would seem to be needed to describe quantum objects; that the random walks of classical objects seem to be characterized by parameters in the sub-Planck domain.

We have found an association of high speed motion with small masses, which, together with large spatial displacements would be compatible with a high rate of diffusion for small masses. At the other extreme of high mass objects, this approach has suggested vanishingly small step speeds for very large masses together with very small spatial displacements, and hence a small rate of diffusion might be associated with large masses. Thus, the diffusion associated with these random walks would seem to be a more important feature of the physics of small masses than large masses; and large mass objects could tend to remain more localized.

Furthermore, in this model, classical behavior would seem to be limited to masses exceeding a threshold mass region comparable to the Planck mass. And it would appear that the behavior of objects with masses below a reference mass near the Planck mass may be expected to be very different from the behavior of classical macroscopic objects of much higher mass.

Together these results point to the possibility of an intrinsic quantum-classical transition which, quantitatively, would seem to be expected in the range of mass near the Planck mass of about 20 micrograms.

In a random walk in ordinary space, an object changes its position by a step at every step in time; whereas, in a random walk in velocity space an object changes its velocity by a step for every step in time. In a physical situation, such a change of velocity would usually correspond to the object's undergoing an impact of some sort, so perhaps it might be preferable to conduct an analysis such as we have conducted here, but in momentum space rather than velocity space, and perhaps it would be useful to attempt to relate the changes in momentum to properties of an interaction causing stochastic impacts. However, in order to avoid going too far afield, we have limited discussion here to the direct characteristics and behavior of the stochastic object itself.

Because the Planck mass has been identified as the constant reference mass that appears in this stochastic analysis, we find that the gravitational constant is present in the parameters of this study. Thus, we see from Eqn. (28) and Eqn. (29) that not only the areal parameter for gravitational acceleration but also the areal parameter for electrostatic acceleration involves the gravitational constant. Thus, it would seem that fundamental physics involving mass and gravitation are covertly present even in electrostatic interactions (perhaps an improved allocation of factors might associate the factor of G with the mass and the stochastic behavior). Furthermore, as just mentioned, it seems possible that the agency responsible for mass and gravitation might also be playing a causative role in the stochastic process.

We have seen that a classical, non-relativistic random walk that would describe ordinary objects would seem to take place in the realm of Planck parameters, and seemingly involve distances even smaller than the Planck length and times even shorter than the Planck time. So what are the characteristics of the realm associated with Planck-scale units which seem to set the range describing this classical stochastic behavior?

On the Planck scale, the concept of a simple, continuous space-time may become inconsistent. By combining concepts from quantum mechanics and general relativity, the existence of fluctuations in space-time on the scale of a Planck length can be anticipated. If gravity is regarded as a field with many of the same properties as the other fundamental force fields, then the state of the gravitational field (and thus the state of space-time itself) would be expected to be, at some level, uncertain and described by quantum physics (although if gravity is exceptional, this would not be the case). Since the general theory of relativity requires that gravitational fields and space-time be the same mathematical objects, space-time itself would also be subject to the kinds of uncertainty required by quantum systems. Because of this, indeterminacy would mean there cannot be precise knowledge of both the geometry of space-time and the rate of change of the space-time geometry, in direct analogy with Heisenberg's uncertainty principle for quantum systems. This indeterminacy for space-time would seem to require that at the Planck scale of about $10^{-35}$ meters and $10^{-43}$ seconds, space-time would exhibit sudden changes in its geometry, and exhibit what has been referred to as a kind of foaminess.[8] So, at that scale, space would no longer be smooth as general relativity defines it. Further theoretical descriptions of space at the scale of the Planck length have been based on loop quantum gravity and superstring theory. In a region the size of the Planck length, the vacuum fluctuations would be expected to be so large that space as we know it becomes what has been described informally as a froth of probabilistic quantum foam or space-time foam, a qualitative description of space-time turbulence at extremely small distances. Thus, the type of stochastic random walk we have been examining would seem to take place in the realm of what has been referred to as quantum foam.[8]

So in this context, one might wonder what actually it means to engage in a random walk that leads to displacements smaller than the Planck length. To complicate the issue, it may be imagined that the properties of the quantum foam are themselves contributing to the symmetric random walk. Can it be that the fluctuations of space-time are the causative agent of the stochastic behavior exhibited as the random walk itself? This simple model that we have been examining appears to be describing a random walk that may be driven by Planck-level physics, even though the Planck-level physics remains in some direct sense invisible. It is interesting that Planck level physics might lead so directly to results consistent with macroscopic phenomena such as Newton's second law and the behavior of classical gravitation and electrostatics.

Or, perhaps the random walk is in some sense invisible or disappears or in some sense ceases to exist as it is lost within the noise of the quantum foam. It is sometimes considered that space and time intervals smaller than the Planck values take us out of "real" space-time. In this context, the result that the stochastic displacements in this random walk of massive classical objects turn out to be so small as to be limited to the sub-Planck domain might be interpreted as indicating that this stochastic motion of classical-sized objects has been made in some sense hidden or invisible by being restricted to displacements within the quantum foam. Such displacements would be literally down in the noise. This would fit in with our sense that quantum-like behavior is largely absent from objects of everyday life, which we regard as exhibiting very well-defined positions and momenta. Can we conclude that classically behaved macroscopic objects are those for which their stochastic behavior in space-time exhibits such small displacements that it is effectively hidden in the Planck-scale quantum foam?

Or, it may simply be considered that an analysis in terms of ordinary space-time at these distances is inherently inappropriate, and that the conclusion should be that there is no "real" random walk in the absence of conventional space-time.

Furthermore, these results would tend to suggest that whatever incompletely understood fundamental physics is responsible for the existence of quantum foam is also causing an intrinsic quantum-classical transition to take place for objects having masses comparable to or larger than the Planck mass value of some 20 micrograms. Might this mass range be somehow determined by a Planck range momentum transfer process?

It should be clear that the present study, although it is employs stochastic ideas, does not specifically address quantum behavior, and so cannot adequately reproduce quantum effects. We are looking here at what appears to be a classical range of stochastic behavior. But what about uncertainties in this model? The stochastic displacement in real position space would seem to provide a measure of spatial uncertainty. The stochastic step displacement in ordinary space $\ell$ is given in Eqn. (27) above, and has the value $2\pi\ell_P m_P^3/m^3 = (h^2/2\pi G)(1/m^3)$. The step displacement $\ell$ might be expected to correspond at least roughly to a spatial uncertainty, $\Delta x$, for the stochastic object, so that we can write:

$$\Delta x = 2\pi\ell_P m_P^3/m^3 \qquad (30)$$

On the other hand, the stochastic displacement in velocity space would be measured by $v_{avg}$, which as estimated in Eqn. (22) would have a value approximately equal to $cm_P^2/m^2 = hc^2/2\pi Gm^2$. This might be presumed to correspond at least roughly with a velocity uncertainty associated with the stochastic object. Thus we could estimate an uncertainty in momentum associated with the stochastic object as the product of the mass and the uncertainty in velocity as:

$$\Delta p = mv_{avg} = cm_P^2/m \qquad (31)$$

From Eqn. (30) and Eqn. (31) together with the definitions of the parameters, we can then find as an estimate of the Heisenberg uncertainty:[11]

$$\Delta p \Delta x = h(m_P^4/m^4) \qquad (32)$$

Accordingly, we see that for masses near the lower limit of validity of this classical approach, that is, for masses comparable to the reference mass or Planck mass $m_P$, that the Heisenberg uncertainty expected would be of the order of Planck's constant h, as might be expected.[11] However, it can also be seen from Eqn. (32) that in the limit of large masses the estimated uncertainty would approach zero, the classical limit. Thus, with respect to the uncertainty principle, while this analysis would lead to a Heisenberg type uncertainty that is comparable to Planck's constant h near the low mass limit of applicability, this uncertainty would go to zero in the high mass classical limit, which although classically acceptable, might seem physically undesirable. However, it appears possible that a form of Heisenberg uncertainty might still be protected even for high masses, since a corresponding quantity formed of Planck units as the product of mass, velocity, and length Planck units ($m_P c \ell_P$) for the quantum foam is equal to Planck's constant $h/2\pi$ and so the classical limit of zero uncertainty might remain effectively invisible, hidden by the uncertainty associated with the quantum foam.

**Summary and conclusions**

We examine a model of a random walk in velocity space, and find that, in the presence of preferential probabilities of steps along a particular direction, such a random walk can lead to an acceleration process resembling that of Newton's second law. An expression for a force-to-mass ratio is obtained.

In three dimensions, this model of a random walk in velocity space leads quite directly to the expectation of inverse square law ($1/r^2$) behavior from forces originating from separated objects located at distances large compared to the distances associated with the stochastic displacements. Thus it seems that some of what we regard as basic characteristics and properties of familiar forces and masses may in a sense be implicit in three dimensional space and stochastic motion.

Both gravitational and electrostatic acceleration are examined, and the relevant stochastic parameters describing the acceleration seem to be compatible with Planck-level parameters. Thus, this simple model appears to be describing a random walk that may be driven by Planck-scale physics. It seems interesting that stochastic motion in conjunction with Planck level physics might lead so directly to macroscopic phenomena such as Newton's second law and classical gravitational behavior.

Furthermore, this model also suggests the possibility of the occurrence of an intrinsic quantum-classical transition in a mass range near the Planck mass of roughly 20 micrograms.

Although this approach greatly oversimplifies, it appears to provide a reasonably straightforward way to model and explore further some nondeterministic aspects of the physical world. Hopefully, this approach may help us experience a different perspective on and perhaps develop a better intuition for what may be taking place in certain aspects of the probabilistic behavior of matter, and hopefully may help to start to develop improved approaches toward some fundamental questions in this area.

Because of the casual approach used in this study, including the various assumptions, conclusions arrived at in this paper must of course be regarded as suggestive and clearly not definitive.

A follow-on paper will address a simplified modeling of a relativistic random walk.